\documentstyle[psfig,prd,preprint,aps]{revtex}
\tightenlines
\newcommand{\be}{\begin{equation}}
\newcommand{\ee}{\end{equation}}
\newcommand{\ba}{\begin{eqnarray}}
\newcommand{\ea}{\end{eqnarray}}
\begin{document}
%%%%%%%%%%%%%%%%%%%%%%%%%%%%%%%%%%%%%%%%%%%%%%%%%%%%%%%%%%%%%%%%%%%
\title{Monopoles, Dyons and Theta Term in Dirac-Born-Infeld Theory}
\author{N.~Grandi\thanks{Becario CICBA}\,, R.L.~Pakman\thanks{Becario CONICET},
F.A.~Schaposnik\thanks{Investigador CICBA} \, and 
G.~Silva$^\dagger$ 
\\
{\normalsize\it
Departamento de F\'\i sica, Universidad Nacional de La Plata}\\
{\normalsize\it
C.C. 67, 1900 La Plata, Argentina}}
 
%\date{\hfill}
\maketitle
 
\begin{abstract}
We present dyon solutions to an $SU(2)$ Dirac-Born-Infeld (DBI)
gauge theory coupled to a Higgs triplet. We consider different
non-Abelian extensions of the DBI action and study the resulting
solutions numerically, comparing them with the standard Julia-Zee
dyons. We discuss the existence of a critical value of $\beta$,
the Born-Infeld absolute field parameter, below which the solution
ceases to exist. We also analyse the effect of modifying 
the DBI action so as to include the analogous of the $\theta$ term, 
showing that Witten formula 
for the dyon charge also holds in DBI theories.
\end{abstract}

\bigskip

%\newpage

\section{Introduction}
The Dirac-Born-Infeld (DBI) action describes the low energy dynamics of 
D-branes \cite{Pol}. In this respect, classical solutions
to the DBI equations of motion have recently received much attention and  
several bion, soliton and instanton configurations have been already
found \cite{G}-\cite{JP}.

Concerning monopole solutions to DBI theory, different possibilities 
have been
discussed, either by considering extensions of the DBI action that
admit BPS equations \cite{NS2} (which are then necessarily the same as in the
Yang-Mills-Higgs system \cite{CNS}) or by coupling the DBI action to
the usual symmetry breaking Higgs Lagrangian  
\cite{gms}-\cite{Pr}. In this last case, 't Hooft-Polyakov-like monopole
solutions arise provided $\beta$, the Born-Infeld ``absolute field'' parameter,
is bigger than a critical value $\beta_c$ \cite{MNS},\cite{gms}.

We extend in the present work the analysis of purely magnetic
solutions of ref.\cite{gms} by constructing electrically charged monopole
solutions to an $SU(2)$ DBI theory coupled to a Higgs field in the adjoint,
with a potential that breaks the symmetry down to $U(1)$. 
Since the  DBI action contains a $(F_{\mu\nu}\tilde F^{\mu\nu})^2$ term
which  vanishes unless electric and magnetic fields are both present,  
dyons test more in detail DBI nonlinearities than purely magnetic monopoles.
Moreover,   one can discuss in this case  some issues concerning Witten effect
\cite{wi1}   and duality \cite{AGZ} in DBI models.

The paper is organized as follows: we start in Section II by discussing
two  alternative ways in which an Abelian DBI action can be extended
to the  case  of an  $SU(2)$ gauge theory depending on the trace
operation used to define a scalar action from non-Abelian fields
\cite{Tse2}. 
Then, we consider the
addition of a Higgs action with a symmetry breaking potential 
and discuss the resulting equations of motion for both trace operations.  
In Section III we discuss dyon configurations by considering 
the usual Julia-Zee
Ansatz \cite{tH}-\cite{JZ}. We find the solutions numerically  
and discuss their main properties. 
The effect of adding a $\theta$-term is studied in section IV where the rules
of charge quantization are discussed in detail.  
Finally,  we present in section V a summary of our results and conclusions.

\section{The Action}

For a non-Abelian gauge group, 
there are alternative definitions of the Dirac-Born-Infeld
action \cite{Tse2},  \cite{AN}-\cite{GSS}. Basically, they 
differ in the way a scalar action is constructed  using different trace 
operations over the group indices. As  
shown by Tseytlin \cite{Tse2}, 
there is one which is singled out by the fact that it leads to an 
action which can be connected to
the tree level string effective action for branes. The corresponding 
Lagrangian is
\be
{\cal L}_{DBI}^{Str}  \equiv \beta^2 {\rm Str}
\left (1-  \sqrt{-{\det} (g_{\mu\nu} + \frac{1}{\beta}  F_{\mu
\nu})} \,\right)  
\label{dbi}
\ee
Here Str is a symmetric trace operation defined by the formula
\be
{\rm Str} (t_1,t_2, \ldots , t_N)
\equiv \frac{1}{N!} \sum_\pi {\rm tr} (t_{\pi(1)} t_{\pi(2)} \ldots
t_{\pi(N)})
\label{5}
\ee
with $t^a$ the generators of the gauge group which we shall take for
simplicity in the fundamental representation of  $SU(2)$, normalized so that
\be
{\rm tr} (t^a t^b) = \delta^{ab}
\label{trace}
\ee
Remarkably, definition (\ref{dbi}) is equivalent to the familiar
Born-Infeld form for the Lagrangian
\be
{\cal L}_{DBI}^{Str} = \beta^2 {\rm Str}  \left( 1 -
\sqrt{1 + \frac{1}{2\beta^2} F_{\mu\nu} F^{\mu \nu}
- \frac{1}{16\beta^4}( F_{\mu \nu}\tilde F^{\mu \nu})^2} \,
\right)\; .
\label{sui}
\ee

In contrast, if one were to use the ``tr'' trace operation 
in the
definition of the DBI action as a determinant like in (\ref{dbi}), 
(of course one has to supplement the definition
with some ordering rule for multiplying determinant elements) its
explicit computation would not lead to the analogous of (\ref{sui}).
One can instead directly  define an alternative DBI Lagrangian 
as
\be
{\cal L}_{DBI}^{tr} = \beta^2 {\rm tr}  \left( 1 -
\sqrt{1 + \frac{1}{2\beta^2} F_{\mu\nu} F^{\mu \nu}
- \frac{1}{16\beta^4}( F_{\mu \nu}\tilde F^{\mu \nu})^2} \,
\right)\; .
\label{sui2}
\ee
We shall then consider both possibilites,  taking as DBI non-Abelian Lagrangian
(\ref{sui}) and (\ref{sui2}).

Apart from this alternative related to the way the trace
operation is defined, one has to decide how the Higgs field
dynamics is introduced.  One possibility is to construct DBI monopoles
 by demanding that the {\it usual}
Yang-Mills-Higgs BPS relations also hold in the DBI case
\cite{NS2}.  This amounts to define a Higgs field Lagrangian  in
a Born-Infeld-like way ({\it i.e.,} with the scalar kinetic
energy and potential terms also under a square root) in
such a way that the model have a  supersymmetric extension
\cite{GGT}, \cite{BP}-\cite{GSS}. One can then prove that the BPS relations 
coincide with those arising
 in the Yang-Mills-Higgs case \cite{CNS}, so that the resulting DBI monopole
solutions are identical to the well-honored Prasad-Sommerfield
exact solutions and have no specific features resulting from the
DBI dynamics. Instead, we shall consider here, as already done in
\cite{gms} for purely magnetic solutions, the usual $SU(2)$
Higgs field Lagrangian and a symmetry breaking potential not
necessarily in the BPS limit. We then propose the following
Lagrangian for the Higgs field:
\be
{\cal L}_{Higgs} =
   \frac{1}{2} D^\mu\vec\phi . D_\mu\vec\phi-V[\phi]
\label{suiss}
\ee
with the scalar triplet  written in the form
\be
\phi = \phi^a t^a = \vec \phi \cdot \vec t \, ,
\label{hi}
\ee
the symmetry breaking potential given by
\begin{equation}
V[\phi] = \frac{\lambda}{4}
\left(\vec\phi \cdot \vec\phi -\frac{\mu^2}{\lambda}
\right)^2  \,
\end{equation}
and the covariant derivative defined as
\be
D_\mu \vec  \phi = \partial_\mu \vec \phi + e \vec A_\mu \wedge
\vec\phi \; .
\label{der}
\ee
Concerning the field strength $F_{\mu \nu} = \vec F_{\mu \nu}. \vec t$,
it is defined as
\be
\vec F_{\mu \nu} = \partial_\mu \vec A_\nu - \partial_\nu \vec A_\mu
+ e\vec A_\mu \wedge \vec A_\nu
\label{f}
\ee
In Born-Infeld theories, it is also convenient to define the
canonically conjugated tensor,
\be
\vec G_{\mu\nu} = -2 \frac{\partial {\cal L}_{DBI}}{\partial \vec 
F^{\mu \nu}}
\label{g}
\ee

\subsection*{(i) The equations of motion  for ${\cal L}^{tr}_{DBI-Higgs}$}
When the usual trace operation ``tr'' is used, the DBI-Higgs Lagrangian
reads
\begin{eqnarray}
{\cal L}^{tr}_{DBI-Higgs}  =  \beta^2 {\rm tr}  \left( 1 -
\sqrt{1 + \frac{1}{2\beta^2} F_{\mu\nu} F^{\mu \nu}
- \frac{1}{16\beta^4}( F_{\mu \nu}\tilde F^{\mu \nu})^2} \right) 
+ \frac{1}{2} D^\mu\vec\phi \cdot D_\mu\vec\phi-V[\phi]\; .
\label{es1}
\end{eqnarray}
 The
equations of motion take the form
\be
D^{\mu}\left(   \frac{\vec{F}_{\mu \nu} 
- \frac{1}{8\beta^2}
\left(\vec F_{\rho \sigma} .\tilde{\vec F}^{\rho \sigma}\right)
\tilde {\vec F}_{\mu \nu} }
{\sqrt{1 + \frac{1}{4\beta^2} \vec F_{\mu\nu}. \vec F^{\mu \nu}
- \frac{1}{64\beta^4}( \vec F_{\mu \nu}. \tilde{\vec F}^{\mu \nu})^2}}
 \right)
  = -e \vec\phi \wedge{ D_{\nu}{\vec \phi}} \; ,
\label{eq1}
\ee
\be
 D^{\mu}D_{\mu} {\vec \phi}=\mu^{2}\vec{\phi} -
\lambda\phi^{2}\vec{\phi}.
\label{eq2}
\ee

\subsection*{(ii) The equations of motion  for ${\cal L}^{Str}_{DBI-Higgs}$}
When the symmetric trace operation is used, the DBI-Higgs Lagrangian is
defined as
\begin{eqnarray}
{\cal L}^{Str}_{DBI-Higgs}  &=&  \beta^2\;\; {\rm Str}  \left( 1 -
\sqrt{1 + \frac{1}{2\beta^2} F_{\mu\nu} F^{\mu \nu}
- \frac{1}{16\beta^4}( F_{\mu \nu}\tilde F^{\mu \nu})^2} \right) +  \frac{1}{2} D^\mu\vec\phi D_\mu\vec\phi-V[\phi] \; .
\label{es2}
\end{eqnarray}
Then, the equations of motion read
\be
D^{ab\,\mu}{\rm Str} \left(   \frac{{F}_{\mu \nu} - \frac{1}{4\beta^2}
\left( F_{\rho \sigma} \tilde{ F}^{\rho \sigma}\right) \tilde { F}_{\mu \nu} }
{\sqrt{1 + \frac{1}{2\beta^2}  F_{\mu\nu}  F^{\mu \nu}
- \frac{1}{16\beta^4}(  F_{\mu \nu} \tilde{ F}^{\mu \nu})^2}}
t^b \right)
  = -e \left(\phi \wedge{ D_{\nu}{\phi}} \right)^a\; ,
\label{Seq1}
\ee
\be
 D^{\mu}D_{\mu} {\vec \phi}=\mu^{2}\vec{\phi} -
\lambda\phi^{2}\vec{\phi}.
\label{Seq2}
\ee

Note that in order to perform the trace operation in the resulting equations
of motion, one has to expand the square root 
and then proceed to the explicit evaluation of traces. While the ``normal''
trace operation tr allows to
reaccomodate the expansion as the square root
appearing in (\ref{eq1}), this is not the case
for the symmetric trace. Then,
  one is left with equations of motion that
  correspond to a $1/\beta^2$
expansion. We just quote here the Lagrangian (\ref{es2}) and 
equations of motion for the gauge field (\ref{Seq1}) expanded
to second 
order in $1/\beta^2$
\begin{eqnarray}
{\cal L}^{(2)} 
&=&
-\frac{1}{4} 
  \vec F_{\mu\nu}
  .
  \vec F^{\mu\nu} 
+
 \frac{1}{192\beta^2}
 \left(
        (
         \vec F_{\mu\nu}
         .
          \vec F^{\mu\nu}
        )^2
      +
        (
          \vec F_{\mu\nu}
          .
          \tilde{\vec F}^{\mu\nu}
       )^2
       +
\right.
\nonumber \\
&&
\!\!\!\!\!\!\!\!\!\!\!\!\!\!\!\!\!\!\!\!\!\!\!\!\!\!\!\!
\left.
         \phantom{ (\vec F_{\mu\nu}.\tilde{\vec F}^{\mu\nu})^2}
         2
         (
          \tilde{\vec F}_{\mu\nu}
          .
          \tilde{\vec F}_{\rho\sigma}
         )
         (
          \vec F^{\mu\nu}
          .
          \vec F^{\rho\sigma}
         )
       +
         2
         (
          \vec F_{\mu\nu}
          .
          \vec F_{\rho\sigma}
         )
         (
          \vec F^{\mu\nu}
          .
          \vec F^{\rho\sigma}
         )
\right)                
\label{simet}
\end{eqnarray}
\begin{equation}
D^\mu \vec G_{\mu\nu} = -e \vec \phi  \wedge D_\nu \vec \phi
\label{Seq3}
\end{equation}
where 
\begin{eqnarray}
\vec G_{\mu\nu}
&=&
  \vec F_{\mu\nu} 
-
  \frac{1}{24\beta^2}
  \left(
          (
           \vec F_{\rho\sigma}
           .
           \vec F^{\rho\sigma}
           )
           \vec F_{\mu\nu}
         +
           (
           \vec F_{\rho\sigma}
           .
           \tilde{\vec F}^{\rho\sigma}
           )
           \tilde{\vec F}_{\mu\nu}          
         +
%\right.
%\nonumber \\
%&&
%\left.
%\phantom{\frac{1}{\beta^2}}
         2(
          \tilde{\vec F}_{\rho\sigma}
          .
          \tilde{\vec F}_{\mu\nu}
          )
          \vec F^{\rho\sigma}
        +
          2
          (
          \vec F_{\rho\sigma}
          .
          \vec F_{\mu\nu}
          )
          \vec F^{\rho\sigma}
\right)
\end{eqnarray}

\section{Dyon solutions}

We   consider the usual spherically symmetric 
Ansatz \cite{tH}-\cite{JZ},
\begin{eqnarray}
&&\vec{A}_{i}(\vec r)= \frac{K(r)-1}{e}\;
\vec{\Omega} \wedge\partial_{i}{\vec{\Omega}}\; , \label{a}\\
&&\vec A_0(\vec r) =  \frac{J(r)}{er}\;
\vec{\Omega}\; , \label{o}\\
&&\vec{\phi}(\vec r)=\frac{H(r)}{er}\;
\vec{\Omega} \; , \label{H}
\end{eqnarray}
with 
\be
\vec \Omega = \vec \Omega(\theta,\varphi) = \frac{\vec r}{r} 
\label{oo}
\ee

The appropriate boundary conditions for $K$, $J$ and $H$,
\be
\lim_{r \to \infty} K(r) = 0 \, , \;\;\;\;\;\;\;\;\; \lim_{r \to 
\infty} \frac{1}{r} J(r) = M   +\frac{b}{r}  \, ,
\;\;\;\;\;\;\;\;\;  \lim_{r \to \infty}
\frac{1}{r} H(r) =
\frac{\mu e}{\sqrt \lambda}
\label{K}
\ee
Here $M$ is a parameter with the dimensions of a mass which has to satisfy
 $M< e\mu/\sqrt\lambda$ to have an appropriate  asymptotic behavior
for $K(r)$ \cite{JZ}. 
Concerning $b$, it determines, as we shall see,
the electric charge.
Concerning  the conditions at the origin, we take
\be
K(0) = 1 \, , \;\;\;\;\;\;\;\;\; J(0) =0 \, , \;\;\;\;\;\;\;\;\;  H(0) = 0 .
\label{HH}
\ee
The electromagnetic $U(1)$ field strength ${\cal F}_{\mu \nu}$
is defined as usual \cite{tH} in the form
\be
{\cal F}_{\mu \nu} = \frac{\vec \phi}{|\vec \phi|}  \cdot 
\left(
\vec F_{\mu \nu}
- \frac{1}{e|\vec \phi|^2}  ( {D_\mu \vec \phi} \wedge
{D_\mu \vec\phi}) \right).
\label{F}
\ee
From (\ref{F}) we define the $U(1)$  
magnetic induction  and electric  field  in the form
\be
B^i = -\frac{1}{2} \varepsilon^{ijk} {\cal F}_{jk} \; ,
\;\;\;\;\;\;\;\;\;\;\;\;\;
E^i = \ {\cal F}^{i0} 
\label{BB-EE}
\ee
Using Ansatz   (\ref{a})-(\ref{oo}) one easily finds that
\be
B^i  = \frac{1}{er^2}\frac{x^i}{r} \, , \;\;\;\;\; E^i = -\left(
\frac{J(r)}{er}
\right)' \frac{x^i}{r}
\label{BBB}
\ee
so that the magnetic flux is
\be
M = \int_{S^2} dS_i  B^i = \frac{4\pi}{e} \, ,
\label{fl}
\ee
It corresponds to that of a unit magnetic monopole located at the
origin. 

Concerning the electric charge, it is   defined as 
\be
Q = \int_{S^2_\infty} dS_i  E^i  =  \frac{4\pi b}{e}
\label{qis}
\ee

In the case of a Born-Infeld theory, it is 
necessary to also define the electromagnetic $U(1)$ 
projection of $G_{\mu \nu}^a$
  which we shall call ${\cal G}_{\mu \nu}$. 
Following the same steps as those leading to ${\cal F}_{\mu \nu}$
(see (\ref{F})), we start from
\be
G_{\mu \nu}^a = {\rm Tr}\left(
 \frac{1}{R}\left(F_{\mu \nu} - \frac{1}{2\beta^2}
(F\tilde F) \tilde F_{\mu\nu} \right) t^a \right) \equiv  {\rm Tr}
\left( 
O_{\mu \nu}^{\rho \sigma} F_{\rho \sigma} t^a \right)
\label{gg}
\ee
Here Tr indicates any one of the two possible trace choices referred
 above and $R$ is defined as
 \be
 R=\sqrt{1 + \frac{1}{2\beta^2} F_{\mu\nu} F^{\mu \nu}
- \frac{1}{16\beta^4}( F_{\mu \nu}\tilde F^{\mu \nu})^2}
 \label{R}
 \ee
Now, inspired in (\ref{F}) we consider the shift 
\be
\vec F_{\mu \nu} \to \vec F_{\mu \nu} - 
\frac{1}{e|\vec\phi|^2}
D_\mu \vec \phi \wedge D_\nu \vec \phi
\label{che}
\ee
in (\ref{gg})
and then project the result on the 
$\check \phi = \vec \phi/|\vec \phi|$ direction. The answer is
\be
{\cal G}_{\mu \nu} = {\rm Tr} \left(
O_{\mu \nu}^{\rho \sigma}\left(
\vec F_{\rho \sigma}  - \frac{1}{e|\vec \phi|^2}
 D_\rho \vec\phi \wedge D_\sigma \vec \phi
\right)\cdot\vec t \frac{\phi}{|\vec \phi|}
\right)
\ee
From ${\cal G}_{\mu \nu}$ we now define the  magnetic
field $H_i$ and the electric induction $D_i$, 
\be
H^i = -\frac{1}{2} \varepsilon^{ijk} {\cal G}_{jk} \; ,
\;\;\;\;\;\;\;\;\;\;\;\;\;
D^i = \ {\cal G}^{i0} 
\label{HH-DD}
\ee

\subsection*{(i) The solution  for ${\cal L}^{tr}_{DBI-Higgs}$} 
Inserting Ansatz (\ref{a})-(\ref{H}) into the eqs. of motion
(\ref{eq1})-(\ref{eq2}) one gets

\begin{eqnarray}
\rho^2 K'' 
&=&
 K
 (K^2 - J^2 + RH^2 - 1)
+
 \frac{\rho ^2R'}{R}
 \left(
         K'
        - 
         \frac{1}{2\hat \beta^2 \rho^3}
        P'JK
 \right) 
 + \frac{\rho}{2\hat\beta^2}
 JK
       \left(
             \frac{P'}{\rho^2}
       \right)'
\nonumber\\
\rho^2 H'' &=& 2H K^2 + \hat \lambda H(H^2 - \rho^2) 
\nonumber
\\
\rho^2J''
&=&
 2JK^2 
+
 \rho
 \frac{R'}{R}
 \left(
        \rho J' - J + \frac{1}{2\hat \beta^2 \rho^2}P'(K^2-1)
        \right)- 
 \frac{\rho}{2\hat \beta^2}
\left(
\frac{P'}{\rho^2}
\right)'
(K^2-1)
 \label{matarr}
\end{eqnarray}
Here we have
\begin{eqnarray}
R^2\! &=&\! 1 + \frac{1}{2\hat\beta^2 \rho^4} 
\!\!\left(
(K^2 - 1)^2 + 2\rho^2K'^2  - 2J^2K^2 - (J-\rho J')^2 
\right)\!\! - \frac{1}{4\hat \beta^4 \rho^4}P'^2\nonumber\\
P &=& \frac{J}{\rho}(K^2-1)
\label{deultima}
\end{eqnarray}
and  we have used dimensionless variables and parameters defined as
\be
  \rho = \frac{e\mu r}{\sqrt{\lambda}} \; ,  \;\;\;\;\;\;\;\;
\hat \lambda = \lambda / e^2 \; , \;\;\;\;\;\;\;\; 
  \hat \beta = \frac{\beta\lambda}{e\mu^2}\; ,  \;\;\;\;\;\;\;\; 
  \hat M = \frac{\sqrt \lambda}{\mu} M 
\label{par}
\ee

Concerning  the energy density one has
\begin{eqnarray}
E &=& \int d^3x \Theta^{00}
=
\frac{4\pi \mu}{e\sqrt{\lambda}}\int d\rho \rho^2 \left(\frac{1}{\rho^4R}
\left( 2 J^2 K^2 +(\rho J' - J)^2 +
        \frac{1}{2\hat \beta^2} P'^2
\right)
+ \right.
\nonumber\\
&& \left.
2\hat \beta^2
(R-1)
+
\frac{\hat \lambda}{4}
\left(
        \frac{H^2}{\rho^2}
      -
        1
\right)^2
+
\frac{1}{2\rho^4}
\left(
        (H - \rho H')^2
       +
         2 H^2 K^2
\right) \right)
\label{energ}
\end{eqnarray}

To obtain a detailed profile of the dyon solutions 
we have solved numerically the differential equations (\ref{matarr})
employing a relaxation method for boundary value problems
\cite{NR}. Such a  method
 determines the solution by starting with an initial guess and
 improving it iteratively. The natural initial guess was
 the exact Prasad-Sommerfield solution \cite{PS} which corresponds
 to $\hat \lambda = 0$ and $\hat \beta \to \infty$.

\begin{figure}%[ht]
\centerline{ \psfig{figure=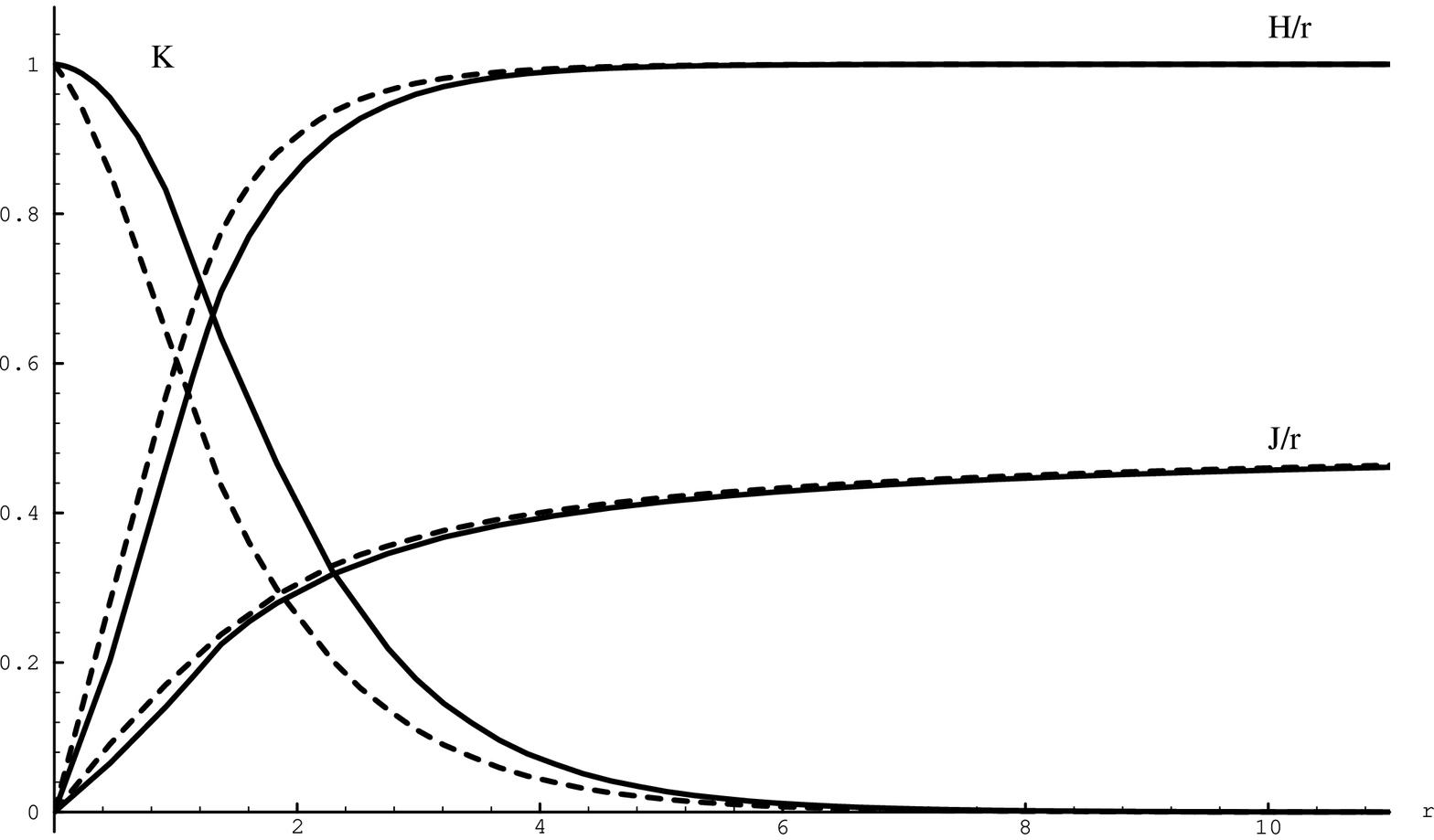,height=20cm,angle=0}}
\smallskip
\caption{ Plot of $K(r)$, $J(r)/r$  and the Higgs field $H(r)/r$ (in
dimensionless variables) for the dyon solution with
$\hat \lambda=0.5$ and $\hat M=0.5$. The solid line corresponds to 
the solution with
$\hat \beta=10$ and the dashed line corresponds to the the solution
with $\hat \beta=0.6$. \label{fig-2} }
\end{figure}

\begin{figure}%[ht]
\centerline{ \psfig{figure=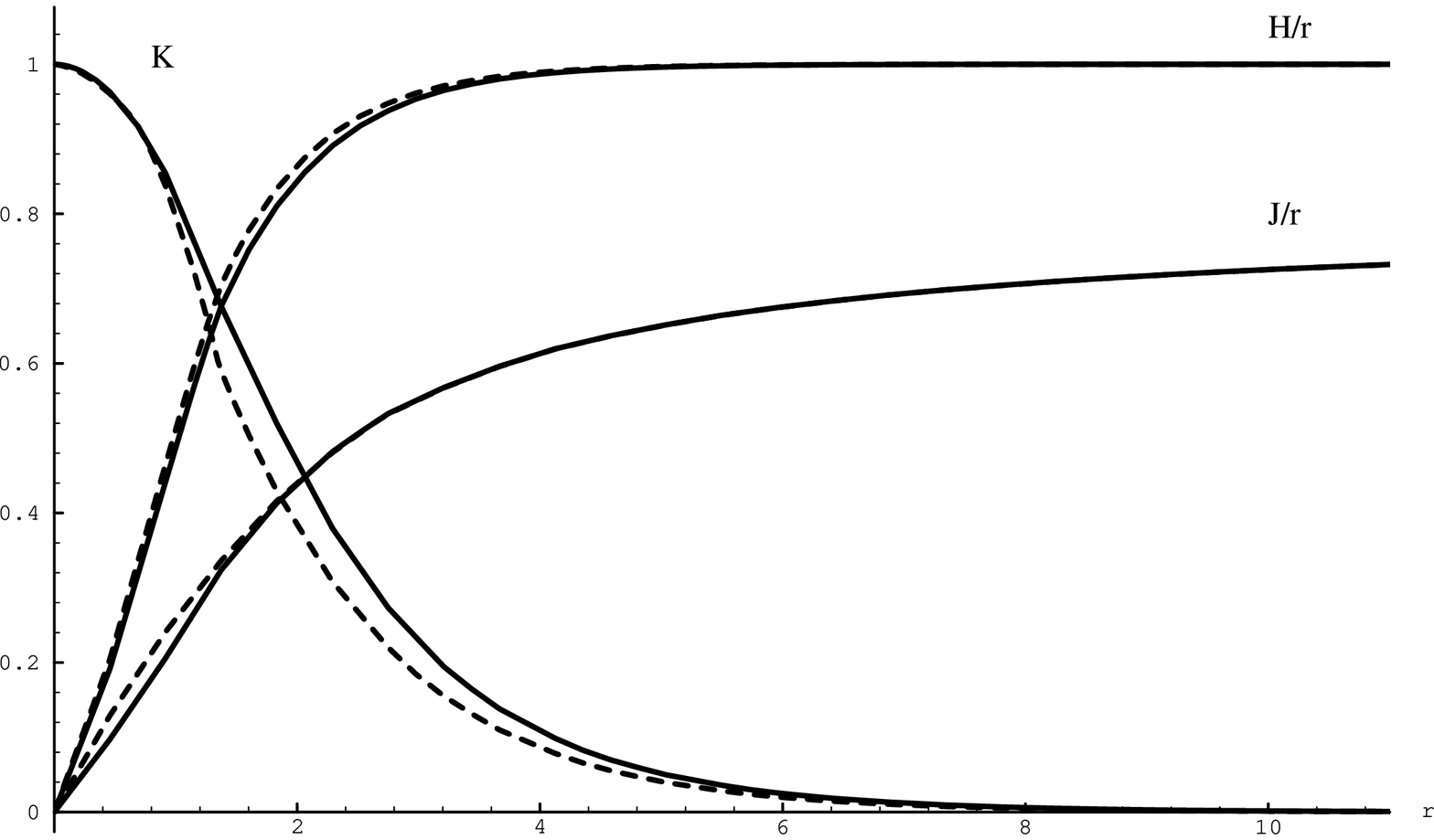,height=20cm,angle=0}}
\smallskip
\caption{ Plot of $K(r)$, $J(r)/r$  and the Higgs field $H(r)/r$ (in
dimensionless variables) for the dyon solution with
$\hat \lambda=0.5$ and $\hat M=0.8$. The solid line corresponds to the solution with
$\hat \beta=10$ and the dashed line corresponds to the the solution
with $\hat \beta=0.6$. \label{fig-3} }
\end{figure}

 For $\hat \beta {\
\lower-1.2pt\vbox{\hbox{\rlap{$>$}\lower5pt\vbox{\hbox{$\sim$}}}}\
} 10$, the solutions to eqs.(\ref{matarr}) do not
differ appreciably from the Julia-Zee dyon  solution
\cite{JZ}. As $\hat \beta$ decreases, the solution changes slowly: the
dyon radius decreases and the (radial) electric and
magnetic fields, $\vec E$ and  $\vec H$ respectively,
concentrate at the origin. Some of the solution profiles are
depicted in figures (\ref{fig-2})-(\ref{fig-3}). It should be
noted that in the limit $\hat M \to 0$ 
we recover the DBI pure monopole
solutions \cite{gms}. 

As it happens for other soliton-like solutions in DBI-Higgs theories
\cite{MNS},\cite{gms}, there is a critical value of $\hat\beta$ which we
call $\hat\beta_c$  such that for $\hat \beta \leq \hat \beta_c$  
the dyon solution
ceases to exist. This can be clearly seen in figure 3, where the energy
is plotted as a function of $\hat \beta$. As $\hat \beta$ 
approaches $\hat \beta_c$ the derivative of the energy with respect
to $\hat \beta$ diverges. We have found that $\hat \beta_c \sim 0.55$
and it does not depend   on $\hat \lambda$ nor on $\hat M$. This yields
a critical dyon radius which is $0.85$ of the standard Yang-Mills
dyon radius (which can be recovered in the $\beta \to \infty$ 
limit) for $\hat M = \hat \lambda = 0.5$. 

The existence of $\hat \beta_c$ is not a byproduct of 
our numerical method but a
genuine effect. As we have thoroughfully discussed in \cite{gms}, the origin
of this phenomenon  could be traced back  to the existence, in DBI theories,
of a second 
dimensionful parameter $\beta$ ($[\beta] = \mu^2$) which enters 
together with $\mu$ in the minimization of the energy.
Indeed, by using approximate solutions,  we have seen in \cite{gms} 
for pure monopoles
(and the same analysis could be done for dyons) that there exist 
a region in the  parameter space defined by the dimensionless combinations
$\hat \lambda$ and $\hat \beta$, for which the energy has no minima. 
This region precisely 
corresponds to small values of $\hat \beta$. 
Of course for the Yang-Mills-Higgs system, where the 
second
dimensionless parameter is absent,  solutions  exist 
in  the whole $\hat \lambda$ range.

One can rephrase the analysis above by noting that when $\hat \beta$ decreases,
the dyon radius also decreases, as can be seen  
in figures 
\ref{fig-2}-\ref{fig-3}. The existence of a critical 
$\hat \beta =\hat \beta_c$ then 
corresponds to the existence of a minimal radius below
which the dyon (or monopole) cannot exist. This is reminiscent of an
analogous
phenomenon that takes place for self-gravitating monopoles
\cite{LNW}-\cite{LW}: they show an instability
for sufficiently strong gravitational coupling,   
manifesting itself as an extremal blackhole in its exterior region
and a more involved solution inside. In other words, 
   non-linearities 
introduced by the DBI action have a similar effect as that produced 
when the coupling to gravity becomes relevant.

\subsection*{(ii) The solution  for ${\cal L}^{Str}_{DBI-Higgs}$} 
As stated above, in order to handle the symmetric trace 
Lagrangian, one has to keep a finite number of terms 
in the expansion of the DBI square root. Then, to order
$1/\beta^2$, one can insert in the eqs. of motion
(\ref{Seq2})-(\ref{Seq3}) the Ansatz \cite{tH}-\cite{JZ}. We shall not display the
 resulting equations but briefly discuss their numerical solution.
 
 For $\hat
 \beta \geq 2$ the solutions  differ less than $ 1 \%$ from
 those arising when the usual trace (``tr'') operation is considered.
 The profile of the solutions are indistinguishable from the 
 solid line curves of figures \ref{fig-2}-\ref{fig-3}. As $\hat \beta$
 decreases, the dyon radius decreases with the same rate as in the usual
 trace case. This signals the existence of a $\hat \beta_c$ also 
 in the symmetric trace case. However,
 since the equations of motion are valid to order $1/\hat \beta^2$, 
 our analysis cannot be reliable for small $\hat \beta$ and the region 
 where one expects 
 to find $\hat \beta_c$ lies outside the validity range of our 
 approximation.

\section{A $\theta$ term}

When  the Yang-Mills-Higgs Lagrangian  includes
a CP violating  $\theta$-term, a remarkable effect 
takes place: a dyon solution  with quantum electric charge $q =n_ee$ and magnetic
charge $g = 4\pi/e$ shifts its electric charge   according
to the relation \cite{wi1}
\be
q = n_ee  + \frac{e\theta}{2\pi}  
\label{ca}
\ee
Here $e$ is the unit electric charge, $n_e$ an integer and a unit magnetic 
charge 
 has been
considered. 
Relation (\ref{ca}) was originally obtained considering an $SU(2)$ gauge theory 
spontaneously broken to $U(1)$ by the vacuum expectation value of a Higgs
triplet, using semiclassical arguments and also by canonical me\-thods.
In this last approach, one defines the operator $N$ that generates 
gauge transformations around the $U(1)$ (electromagnetic) surviving symmetry
and then imposes as an operator statement
\be
\exp(2\pi i N) = I
\label{ent}
\ee
Now, when the Lagrangian includes a CP violating term of the form
\be
\Delta L = \theta \frac{e^2}{32\pi^2}{\rm tr}\tilde F_{\mu\nu}F^{\mu \nu}
\label{del}
\ee
one can see that the condition (\ref{ent})
implies formula (\ref{ca}).  For a general dyon solution, one gets
\be
q = n_e e + \frac{e\theta }{2\pi} n_m
\label{general}
\ee
with the magnetic charge    $g$  expressed as a multiple of the unit 't Hooft-Polyakov charge,
\be
g = \frac{4\pi}{e} n_m  
\label{gene}
\ee

In this section we analize whether 
a similar phenomenon can take place in the non-Abelian Dirac-Born-Infeld theory
we have described above.  
 
Let us first recall
that in Yang-Mills theory, a $\theta$-term can be generated
 by the action of an $SO(2)$ rotation followed
by a scaling of the field strength \cite{GR}.  Indeed, if one considers
\begin{equation}
F_{\mu \nu} \to 
\frac{1}{\sqrt{\cos 2\alpha}}
\left(\cos \alpha F_{\mu \nu} - \sin \alpha \tilde F_{\mu \nu}
\right) 
\label{shift1}
\end{equation}
then, a $\theta$ term of the form (\ref{del}) can be generated from a
$(-1/4) {\rm tr} F_{\mu \nu}F^{\mu \nu}$ term, with $\alpha$ related to $\theta$
through the formula
\be
\tan(2 \alpha) = \frac{e^2\theta}{8\pi^2}
\label{teta}
\ee

We shall now see that the same transformation changes the Dirac-Born-Infeld 
Lagrangian in such a way that when one computes the electric charge of dyons,
  Witten effect takes place exactly as   in Yang-Mills theory.
  
We  start then by analysing the effect of transformations 
(\ref{shift1})  in the
DBI action. For definiteness, we shall consider the case in which
the DBI action is defined using the symmetric trace. Then, 
performing  the change (\ref{shift1}) in (\ref{dbi}), one gets
\be
{\cal L}^{\theta}_{DBI}  = \beta^2 {\rm Str}
\left (1-  \sqrt{-{\det} 
\left(g_{\mu\nu} +  
\frac{1}{\beta\sqrt{\cos 2\alpha}}\left(
\cos \alpha F_{\mu \nu} - \sin \alpha \tilde F_{\mu \nu}
\right)
\right)} \,\right)
\label{dbit}
\ee
which can be also written in the   form

\begin{eqnarray}
& & \!\!\!\!
{\cal L}_{\theta}  =  
\beta^2 {\rm Str}  \left( 
\phantom{\sqrt{\sum_{N=1}^N}}
\!\!\!\!
\!\!\!\!\!\!\!\!\!
1 -
\right.\nonumber\\
& & \left.
\sqrt{1 + \frac{1}{2\beta^2} 
\left(
F_{\mu\nu} F^{\mu \nu} - \frac{e^2\theta}{8\pi^2}\tilde
F_{\mu\nu} F^{\mu \nu}
\right) 
- \frac{1}{16\beta^4}
\left( \tilde F_{\mu \nu} F^{\mu \nu} +
\frac{e^2\theta}{8\pi^2}
F_{\mu\nu} F^{\mu \nu}
\right)^2} \,
\right)
\nonumber\\
\label{suit}
\end{eqnarray}
%

%%%%%%%%%%%%%%%%%%%%%%%%%
In contrast
with the case of YM theory, where the addition of a $\theta$-term 
does not change the eqs. of motion ($\tilde F F$ is a surface term), here, 
rotation (\ref{shift1})
leads to eqs. of motion that differ from the $\theta = 0$ ones. Then,
one has to see whether   dyon solutions still exist for $\theta \ne 0$. 
Now, studying the modified system of eqs. of motion 
to the order   worked out in the $\theta = 0$ case (i.e., up to
the order $1/\hat \beta^2$) one can easily see that its solutions coincide 
(if one rescales $\hat \beta$ conveniently) with the $\theta=0$ ones.

Having found that dyon solutions exist when a $\theta$ term is present, we
are now ready 
to explicitly write  the operator $N$ that generates the transformations
associated with the surviving symmetry. One
 has to consider
$U(1)$ transformations along the Higgs field direction $\check\phi$,
\begin{eqnarray}
\delta_{U(1)} \phi &=& 0\nonumber\\
\delta_{U(1)} A_\mu &=& \frac{1}{e}D_\mu (\epsilon \check \phi)
\label{suich}
\end{eqnarray}
Then, the corresponding
conserved current picks a  contribution solely from the DBI Lagrangian,  
\be
J_\mu^{U(1)} =   
\frac{\partial {\cal L}^\theta_{DBI}}{\partial\left(\partial^{\mu}A^{a\,\nu}
\right)} \delta_{U(1)} A^{a\,\nu}
\ee
and the conserved charge $N$ then takes the form
\be
N = \int d^3x J_0^{U(1)} = 
\frac{1}{e}\int d^3x \partial^i \left( G^a_{0i} 
\check\phi^a\right) - \frac{1}{e}\int d^3x  \check\phi^a \left(D^iG_{0i}\right)^a
\label{N}
\ee
where
\be 
G_{\mu \nu}^a  
= {\rm Str}\left(t^a
\frac{1}{R_\theta} \left(
F_{\mu \nu} - \frac{e^2\theta}{8\pi^2}
 \tilde F_{\mu \nu} - \frac{1}{4\beta^2}
(F_{\rho \sigma}\tilde F^{\rho \sigma} + \frac{e^2\theta}{8\pi^2} 
F_{\rho \sigma}F^{\rho \sigma})
  (\tilde F_{\mu \nu} +
 \frac{e^2\theta}{8\pi^2} F_{\mu \nu})
 \right) \right)
\label{G}
\ee
with
\be
R_\theta = \sqrt{1 + \frac{1}{2\beta^2} 
\left(
F_{\mu\nu} F^{\mu \nu} - \frac{e^2\theta}{8\pi^2}\tilde
F_{\mu\nu} F^{\mu \nu}
\right) 
- \frac{1}{16\beta^4}
\left( \tilde F_{\mu \nu} F^{\mu \nu} +
\frac{e^2\theta}{8\pi^2}
F_{\mu\nu} F^{\mu \nu}
\right)^2}
\label{RF}
\ee

It is important to note that
\be
\lim_{r \to \infty}  G_{\mu \nu} = F_{\mu \nu} - \frac{e^2\theta}{8\pi^2}
\tilde F_{\mu \nu}
\label{rela}
\ee

Now, use of
the  equations of motion makes the second term in the r.h.s.
of (\ref{N})  vanish. Then,
\be
eN =  \int_{S^2_\infty}dS^i G^a_{0i}  \check\phi^a  =
\int _{S^2_\infty}dS^i \left(
F^a_{0i}   - \frac{e^2\theta}{8\pi^2}
\tilde F^a_{0 i}
\right)\check\phi^a 
\label{otra}
\ee
or, using the magnetic charge $M$  and electric charge $Q$ defined by eqs.
(\ref{fl}) and (\ref{qis}),
\be
eN = Q - \frac{\theta e^2}{8 \pi^2} M
\label{wif}
\ee
Condition (\ref{ent}) implies that $N$ has to have integer eigenvalues $n_e$. 
Calling  $q$ and $g$ the eigenvalues of the electric and magnetic charge 
operators $Q$ and $M$ respectively, we then have
\be
q = e n_e +\frac{\theta e^2}{8 \pi^2} g
\label{devue}
\ee
Now, we have seen 
that the DBI theory with a  $\theta$ term admits monopole solutions
  with  unit magnetic charge $ g = (4\pi/e)$ so that formula (\ref{devue}) coincides with (\ref{general}) obtained for the Georgi-Glashow
model if one considers a solution with $n_m$ units of magnetic charge.  We then conclude that for the DBI model with    $\theta$ term
the basic formulae  (\ref{general}) and (\ref{gene}) hold. 
One can then introduce the complex parameter $\tau$ \cite{AGZ} and, from the
resulting discrete two-dimensional lattice, infer the existence of a discrete
$SL(2,Z)$ symmetry. Of course to thoroughfully study electric-magnetic duality,
one should at this point consider the supersymmetric extension of DBI models
but this goes beyond the scope of the present investigation. 

\section{Summary and Conclusions}

We have seen that spontaneously broken Dirac-Born-Infeld  $SU(2)$ gauge 
theory admits dyon solutions which, in the range 
$\hat \beta > \hat \beta_c$ behave as Julia-Zee dyons in Yang-Mills theory.
As the absolute field parameter $\hat \beta$ decreases, the radius of the
dyon also decreases so that the magnetic and electric field become 
more and more concentrated. For $\hat \beta \leq \hat \beta_c$ we have seen
that the dyon solution ceases to exist, much in the way self-gravitating 
monopole and dyon solutions become unstable when coupling to 
gravity is sufficiently strong: in both cases there is a minimum radius
below which the solution collapses.

Once the existence of monopole and dyon solutions is proven, it is natural to
consider whether the analogue of Witten effect takes place in theories in 
which the gauge field dynamics is dictated by a DBI action. To study this 
issue,  one has to include a theta term which, in the present case, arises 
naturally after an $SO(2)$ rotation in $F_{\mu \nu}$ is performed. Remarkably,
athough this shift greatly complicates the DBI dynamics, one can prove, using
the Noether method, that the dyon electric charge is shifted exactly
in the same way as in the Yang-Mills case. This makes natural to study the
issue of duality in the supersymmetric extension of DBI theory \cite{GSS},
\cite{CNS}. We hope to discuss this problem in a future work.

~

\underline{Acknowledgements}: We wish to thank Eduardo Fradkin
for a helpful comment.
F.~A.~S. is
partially  supported as Investigador by CICBA (Argentina). R.~L.~P. wishes 
to thank CICBA (Argentina) for support in the first stages of this work.
This work is partially supported  by CONICET and  ANPCYT (PICT 97/2285) 
(Argentina).

%\newpage

\newpage
%
%%%%%%%%%%%%%%%%%%%%%%%%%%%%%%%%%%%%%%%%%%%%%%%%%%%%%%%%%%%%%%%%%

%%%%%%%%%%%%%%%%%%%%%%%%%%%%%%%%%%%%%%%%%%%%%%%%%%%%%%%%%%%%%%%%%
% Figure 1
%%%%%%%%%%%%%%%%%%%%%%%%%%%%%%%%%%%%%%%%%%%%%%%%%%%%%%%%%%%%%%%%%

% Figure 2
%%%%%%%%%%%%%%%%%%%%%%%%%%%%%%%%%%%%%%%%%%%%%%%%%%%%%%%%%%%%%%%%%

%%%%%%%%%%%%%%%%%%%%%%%%%%%%%%%%%%%%%%%%%%%%%%%%%%%%%%%%%%%%%%%%%
% Figure 4
\begin{figure}%[ht]
\centerline{ \psfig{figure=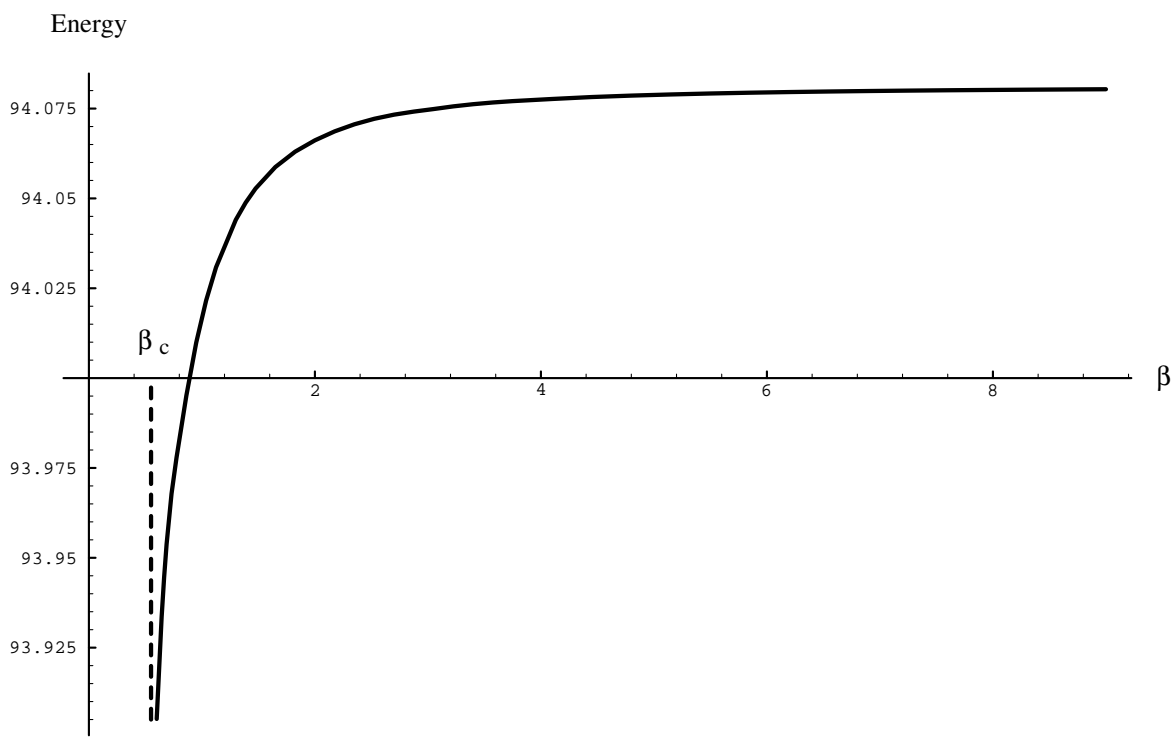,height=20cm,angle=0}}
\smallskip
\caption{ Energy of the dyon configuration as a function of
$\hat \beta$ for $\hat M = 0.8$ and  $\hat \lambda = 0.5$ 
(Similar curves are obtained for other values of $\hat M$ and  
$\hat \lambda$). \label{fig-4} }
\end{figure}
%%%%%%%%%%%%%%%%%%%%%%%%%%%%%%%%%%%%%%%%%%%%%%%%%%%%%%%%%%%%%%%%%

\end{document}